\documentclass[11p]{article}
\usepackage[dvipdfmx]{graphicx}
\usepackage{amsmath}
\usepackage{color}
\usepackage{tikz}
\usepackage{ulem}
\usepackage{caption}

\newtheorem{proposition}{Proposition}[section]
\newtheorem{theorem}{Theorem}[section]
\author{
HIDETOMO NAGAI$^{\dag }$\\
\textit{Department of Mathematics, Tokai University, 4-1-1,} \\
\textit{Kitakaname, Hiratsuka, Kanagawa, 259-1292, Japan}  \\
$^{\dag }$Corresponding author.  Email:hdnagai@tokai-u.jp \\
AND \\
NOBUHIKO SHINZAWA \\
\textit{Department of Engineering, Nishinippon Institute of Technology,} \\
\textit{1-11, Aratsu, Kandamachi, Miyako-gun, Fukuoka, Japan}\\
}
\title{\textbf{Discrete and Ultradiscrete Mixed Soliton Solutions}}

\date{} 

\begin{document}

\maketitle

\begin{abstract}
We propose a new type of soliton equation, which is obtained from the generalized discrete BKP equation.  The obtained equation admits two types of soliton solutions.  The signs of amplitude and velocity of the soliton solution are opposite to the other.  We also propose the ultradiscrete analogues of them.  The ultradiscrete equation also admits the similar properties.  In particular it behaves the original Box-Ball system in a special case.  \end{abstract}
\section{Introduction}
Soliton equations are known as nonlinear equations which have exact solutions.  
Among these equations, the soliton equations are classified in some hierarchies.  For example, the KP equation, BKP equation\cite{DKP}.  In the KP equation, the KdV equation or the Toda equation have been researched especially.  This is because they have good properties and to easy to handle.  Similarly, the Sawada-Kotera equation is also known as a good example in the BKP equation.  \par
Soliton equations can be divided into continuous, discrete and ultradiscrete equations by its discreteness of dependent and independent variables.  In discrete soliton equations, there are also the discrete KP, BKP equations similar to the continuous equations.  Recently, one of the authors proposed the generalized discrete BKP equation and its soliton solution\cite{Shinzawa}.  The equation is expressed by 
\begin{equation}\label{eq:gdBKP}
\begin{aligned}
  &z_1\tau(p+1, q, r)\tau(p, q+1, r+1)+z_2\tau(p, q+1, r)\tau(p+1, q, r+1)\\
  &+z_3\tau(p, q, r+1)\tau(p+1, q+1, r)-z_4\tau(p, q, r)\tau(p+1, q+1, r+1) =0, 
\end{aligned}
\end{equation}
where $z_1$, $z_2$, $z_3$, $z_4$ are arbitrary parameters satisfying $z_1+z_2+z_3-z_4=0$.  We note (\ref{eq:gdBKP}) reduces to the discrete KP equation when $z_4=0$.  Moreover, when we set 
\begin{equation}  \label{BKP}
\begin{aligned}
  &z_1=(a+b)(a+c)(b-c), \quad z_2=(b+c)(b+a)(c-a), \\
  &z_3=(c+a)(c+b)(a-b), \quad z_4=-(a-b)(b-c)(c-a),  
\end{aligned}
\end{equation}
then (\ref{eq:gdBKP}) reduces to the discrete BKP equation\cite{Miwa}. 
This fact means the generalized discrete BKP equation includes the discrete KP, BKP equations.  
Conversely, as mentioned in \cite{Shinzawa}, (\ref{BKP}) have solutions for $a$, $b$, $c$ when $z_1z_2z_3z_4\not =0$.  In other words, the generalized discrete BKP equation (\ref{eq:gdBKP}) coincides with the discrete  BKP equation for most values of $z_i$.  However, the advantage of adopting the expression of (\ref{eq:gdBKP}) than that of the discrete BKP equation is the coefficients can be taken freely.  
This freedom gives us a simple expression of soliton solutions.  Furthermore it enables us to ultradiscretize equations and solutions easily.  \par 
In this paper we  present a new type of soliton equation from (\ref{eq:gdBKP}) through the reduction.  
The obtained equation possesses exact solution which contains two different types of soliton solutions.  The signs of amplitudes and the velocities are opposite each other.  We also give its ultradiscrete analogues\cite{Tokihiro}.  The obtained ultradiscrete solution also contains two different types of solutions.  One behaves as the original Box-Ball system and the other does negative soliton solution. \par 
This paper composed below.   In Section 2, we derive a certain discrete soliton equation from (\ref{eq:gdBKP}).  Also we give the ultradiscrete equation.  In Section 3, we give the discrete soliton solutions and show its properties.  In Section 4, we ultradiscretize the discrete soliton solutions.  In Section 5, we take the continuous limit and show the relation between our equation and the KdV-Sawada-Kotera equation.  We give concluding remarks in Section 6.   
\section{Discrete and Ultradiscrete Equations}
In this Section, we derive new discrete and ultradiscrete equations from (\ref{eq:gdBKP}) by imposing some conditions.  First we introduce new independent variable transformations  
\begin{equation}  \label{trans}
  p=n-m-2l, \quad q=-l, \quad r=m+l.   
\end{equation}
Then (\ref{eq:gdBKP}) is rewritten by  
\begin{equation}  \label{eq:gdBKP2}
\begin{aligned}
 &z_1\hat \tau(l, m+1, n)\hat \tau(l+1, m-1, n+1)+z_2\hat \tau(l, m, n-1)\hat \tau(l+1, m, n+2)\\
 & +z_3\hat \tau(l, m, n)\hat \tau(l+1, m, n+1)-z_4\hat \tau(l+1, m-1, n)\hat \tau (l, m+1, n+1)=0,    
\end{aligned}
\end{equation}
where $\hat \tau $ depends on $l$, $m$, $n$.  In particular, we set 
\begin{equation} \label{def:z}
z_1=d_1, \qquad z_2=d_2, \qquad z_3=1-d_2, \qquad z_4=1+d_1, 
\end{equation}
with $0< d_2<d_1<1$.  By imposing $\hat \tau $ does not depend on $l$, that is, $\hat \tau(l+1, m, n)=\hat \tau(l, m, n) $ holds, then we finally obtain 
\begin{equation}  \label{eq:dmix}
  (1+d_1)f^{m-1}_{n}f^{m+1}_{n+1} =d_1f^{m+1}_{n}f^{m-1}_{n+1}+d_2f^m_{n-1}f^m_{n+2}+(1-d_2)f^m_nf^m_{n+1},  
\end{equation}
where $f^m_n$ denotes $\hat \tau (l, m, n)$.  We note (\ref{eq:dmix}) is reduced to the discrete KdV equation in bilinear form when $d_2=0$\cite{Tsujimoto}.  
In addition we note (\ref{eq:dmix}) can be also obtained from the discrete BKP equation by putting suitable parameters $a, b, c \in \bf{C}$ into (\ref{BKP}), however the expressions of them are complicated to deal with.  \par
If we introduce dependent variable transformations 
\begin{equation}  \label{def:uvx}
  u_n^m=\frac{f_{n+1}^mf_n^{m+1}}{f_n^mf_{n+1}^{m+1}}, \qquad   v^m_n = \frac{f_{n}^{m+1}f_n^{m-1}}{(f_{n}^{m})^2}, \qquad  x^m_n = \frac{f_{n+1}^{m}f_{n-1}^{m}}{(f_{n}^{m})^2},   
\end{equation}
for (\ref{eq:dmix}), we obtain 
\begin{subequations} \label{eq:dnon}
\begin{align}  
  & \frac{u^{m}_{n}}{u^{m-1}_{n}}=  \frac{v^{m}_{n}}{v^{m}_{n+1}} , \label{eq:dnonu} \\
  &(1+d_1) v^m_{n+1}=(1-d_2)u^{m-1}_n+d_1v^m_n(u^{m-1}_n)^2+d_2x^m_{n+1}x^m_n u^{m-1}_n  ,\label{eq:dnonv} \\
  &\frac{x^{m}_{n}}{x^{m-1}_{n}}=  \frac{u^{m-1}_{n-1}}{u^{m-1}_{n}} . \label{eq:dnonx}
\end{align}
\end{subequations}
We regard $m$ and $n$ as time and space variables.  Then the time evolutions of $u^m_n$, $v^m_n$, $x^m_n$ are described by (\ref{eq:dnonu}), (\ref{eq:dnonv}) and (\ref{eq:dnonx}) under the boundary condition $\lim_{n\to -\infty} v^m_{n}=const. $  \par   
Next, we shall derive the ultradiscrete analogues\cite{Tokihiro}.  By introducing transformations $d_i=e^{-\delta _i/\varepsilon }$, $f_n^m=e^{F_n^m/\varepsilon }$, and taking the limit $\varepsilon \to +0$ for (\ref{eq:dmix}), we obtain the ultradiscrete equation.  
\begin{equation}  \label{eBB}
 F_{n+1}^{m+1}+F_{n}^{m-1}=\max( F_{n+1}^{m}+F_{n}^{m}, F_{n}^{m+1}+F_{n+1}^{m-1}-\delta _1, F_{n+2}^{m}+F_{n-1}^{m}-\delta _2),   
\end{equation}
where $0< \delta _1< \delta _2$.  Here we use the key formula $\lim_{\varepsilon \to +0}\varepsilon \log (e^{A/\varepsilon}+e^{B/\varepsilon})=\max(A, B)$.  We note (\ref{eBB}) reduces to the ultradiscrete KdV equation in bilinear form when $\delta_2 $ is sufficiently large\cite{uKdV}.  
We can also obtain the ultradiscrete analogues of (\ref{eq:dnon}) with the transformations $u_n^m=e^{U_n^m/\varepsilon }$, $v_n^m=e^{V_n^m/\varepsilon }$, $x_n^m=e^{X_n^m/\varepsilon }$ by the similar procedure.  
\begin{subequations}
\label{BBBshin}
\begin{align}
U^m_n =& U^{m-1}_n + V^m_n - V^m_{n+1}, 
\label{BBB1shin} \\ 
V^m_{n+1} =& \max(0,  V^m_n + U^{m-1}_n-\delta_1, X^m_{n+1} + X^m_n- \delta_2) + U^{m-1}_{n}, 
\label{BBB2shin} \\
X^{m}_n =& X^{m-1}_n+U^{m-1}_{n-1} - U^{m-1}_n.  
\label{BBB3shin}
\end{align}
\end{subequations}
Notice that the equations (\ref{BBB1shin}) and (\ref{BBB3shin}) take the 
form of the conservation law. 
This procedure determines $X^m_{n}$ from the values of $X^{m-1}_n$ and $U^{m-1}_n$ from (\ref{BBB3shin}).
Then, if we assume boundary condition $\lim _{n\to -\infty}V^m_n$ we can determine the values of $V^{m}_{n+1}$ from the values of $U^{m-1}_n$, $V^{m}_n$, $X^{m}_n$ using  (\ref{BBB2shin}).  Finally $U^m_n$ is determined from the values of $U^{m-1}_n$ and $V^m_n$ from (\ref{BBB1shin}).   
Repeating this procedure to the new values $U^{m}_n$, $X^{m}_n$, we can obtain the time evolution.

Eq.(\ref{BBBshin}) can be considered as a kind of Box-Ball systems.
Let us consider $U^m_n$ and $V^m_n$ as a number of balls in $n$th box at time $m$ and a carrier of balls respectively.  Also let us consider $\delta _1$ as the capacity of the boxes.  Assume the boundary conditions 
\begin{eqnarray*}
\lim_{n \rightarrow - \infty} U^m_n
= \lim_{n \rightarrow - \infty} V^m_n
= \lim_{n \rightarrow - \infty} X^m_n = 0.    
\end{eqnarray*}
Then the time evolution rule of (\ref{BBBshin}) is described as follows.  
From time $m-1$ to $m$, the carrier moves from the $-\infty$ site to the $\infty$ site.  It passes each box from the left to the right.  At $n$th box, the carrier loads the balls as many as possible from the $n$th box and at the same time, unloads the balls from the carrier as many as the vacant spaces of the $n$th box, which corresponds to $\delta _1- U^{m-1}_n$.  Moreover, if the maximum value of RHS in (\ref{BBB2shin}) is given by $X^m_{n+1}-X^m_n-\delta _2$, then the carrier loads more balls even if the number of balls becomes negative.  In this case we consider the box has `negative balls' as a debt.  

As a special case, if the condition $-\delta_2 + X^m_n +  X^m_{n+1} < 0$ always 
holds, then (\ref{BBB2shin}) becomes 
\begin{eqnarray}
V^m_{n+1} = \max(0, V^m_n + U^{m-1}_n- \delta_1) + U^{m-1}_{n}.  
\label{BB1shin}
\end{eqnarray}
Using the conservation law (\ref{BBB1shin}) 
to the left hand side of this equation, we obtain the following.
\begin{eqnarray}
U^{m}_n = \min (V^m_n,\ \delta_1 - U^{m-1}_n).   \label{BB2shin}
\end{eqnarray}
We can also rewrite the dependent variable $V^m_n$ as 
\begin{equation}
V^m_n = \sum_{k= - \infty}^{n}(V^m_{k} - V^m_{k-1}) 
= \sum_{k=-\infty}^{n} (U^{m-1}_{k-1} - U^{m}_{k-1}), 
\end{equation}
using the conservation law (\ref{BBB1shin}) 
together with the boundary condition $\displaystyle \lim_{n \rightarrow -  \infty} V_n^m = 0$.
Substituting this representation to the equation (\ref{BB2shin}), 
the equations (\ref{BB1shin}) and the conservation law (\ref{BBB1shin}) 
turns out to be 
\begin{equation}
U^{m}_n = \min \left(\sum_{k=-\infty}^{n} (U^{m-1}_{k-1} - U^{m}_{k-1}) ,\ \delta_1 - 
	    U^{m-1}_n \right). \label{BBshin2}
\end{equation}
This is nothing but the Box-Ball system with the box capacity $\delta_1$.  \par
Exact solutions and examples for (\ref{BBBshin}) are given in Section 4.  
\section{Discrete Soliton Solution} 
The soliton solution of (\ref{eq:gdBKP}) is expressed by
\begin{equation}  \label{Nsol0}
  \tau(p, q, r)= \sum _{0\le k\le N} \sum _{I_k \subset [N] } \left( \prod _{i\in I_k} c_i\frac{\varphi(t_i)}{\varphi(s_i)} \prod _{i, j\in I_k \atop{i<j }} b_{ij}\right),
\end{equation}
where $N$ is a positive integer, $[N]$ denotes $\{1, 2, \dots , N \}$ and $\sum _{I_k \subset [N] }$ the summation over all $k$-element subsets $I_k=\{ i_1, i_2, \dots , i_k\}$ chosen from $[N]$.  When $k=0$, we define $\sum _{I_0\subset [N]}$ is $1$.  Function $\varphi (t)$  and $b_{ij}$ are defined by  
\begin{equation}  \label{def}
\begin{aligned}
  &\varphi(t) = (a_1(t))^p (a_2(t))^q (a_3(t) )^r, \\
  & a_1(t)= \frac{z_2z_4+z_3t}{z_1-t}, \qquad  a_2(t)= \frac{z_1z_4-z_3t}{z_2+t}, \qquad  a_3(t)= t, 
\end{aligned}
\end{equation}
\begin{equation}  \label{def of b}
\begin{aligned}
  &b_{ij} = \frac{cf(t_i, t_j)cf(s_i, s_j)}{cf(s_i, t_j)cf(t_i, s_j)}  , \qquad  cf (t, s) = \frac{t-s}{tsz_3+z_1z_2z_4}.  
\end{aligned}
\end{equation}
Note the solution (\ref{Nsol0}) is derived from the original one given in \cite{Shinzawa} through the gauge transformation(See Appendix).  
First we shall consider the conditions for parameters $t_i$ and $s_i$ so that $\tau(p, q, r)$ take positive values for any $p$, $q$, $r$.  Hereafter we suppose $z_1$, $z_2$, $z_3$, $z_4$ are positive.  The following proposition holds.  
\begin{proposition}  \label{prop:2-1}
  Set $t_i'=-\frac{z_1z_2z_4}{z_3s_i}$, $s_i'=-\frac{z_1z_2z_4}{z_3t_i}$.  Then 
\begin{equation}  \label{rel:prop2-1}
  \frac{\varphi(t_i)}{\varphi(s_i)}=\frac{\varphi(t'_i)}{\varphi(s'_i)} , \qquad b_{ij}=b_{i'j}=b_{ij'}=b_{i'j'}
\end{equation}
hold for any $i, j=1, 2, \dots , N$ $(i\not= j)$.  Here $b_{i'j}$ denotes
\begin{equation}
  b_{i'j} = \frac{cf(t'_i, t_j)cf(s'_i, s_j)}{cf(s'_i, t_j)cf(t'_i, s_j)}.  
\end{equation}
\end{proposition}
\textbf{Proof.} \ The relations (\ref{rel:prop2-1}) can be obtained by confirming
\begin{equation}
  \frac{a_k(t)}{a_k(s)}=\frac{a_k(t')}{a_k(s')}, \qquad cf(t_i, t_j)  =-\frac{1}{z_1z_2z_3z_4 cf(t_i, s_j')} = cf(s_i', s_j')  
\end{equation}
for $k=1, 2, 3$ from the definitions.  \par
From Proposition \ref{prop:2-1}, we may assume either $t_i$ or $s_i$ is always positive for any $i$ without loss of generality.  Actually, the solution with $t_i$, $s_i$ and the one with $t_i'$, $s_i'$ correspond.  Moreover, considering the term $ (a_3(t_i)/a_3(s_i))^r=(t_i/s_i)^r$ in $\varphi(t_i)/\varphi(s_i)$, we have the condition both $t_i$ and $s_i$ should be positive so that $\varphi(t_i)/\varphi(s_i)>0$.  Similarly, by considering the positivity of $a_1(t_i)/a_1(s_i)$, $a_2(t_i)/a_2(s_i)$, we obtain the following proposition.  
\begin{proposition}  \label{prop:2-2}
If parameters $t$, $s$ satisfy one of the conditions,
\begin{enumerate}
  \item $0< t, s< \underline{z}$,
  \item $\underline{z}< t, s< \overline{z}$,
  \item $\overline{z}< t, s$,
\end{enumerate}
where $\underline{z}$ and $\overline{z}$ denote  $\min(z_1, \frac{z_1z_4}{z_3})$, $\max(z_1, \frac{z_1z_4}{z_3})$ respectively, then $\varphi(t)/\varphi(s)$ take positive values for any $p, q, r\in \bf{Z}$. 
\end{proposition}
We also have the following.  
\begin{proposition}  \label{prop:2-3}
  Suppose $t_i$, $s_i$ are positive for any $i=1, 2, \dots, N$.  If parameters $t_i$, $s_i$ satisfy one of the conditions, 
\begin{enumerate}
  \item Both $t_i$ and $s_i$ take values between $t_j$ and $s_j$,
  \item Both $t_j$ and $s_j$ take values between $t_i$ and $s_i$,
  \item $\max(t_i, s_i) < \min(t_j, s_j)$, 
  \item $\max(t_j, s_j) < \min(t_i, s_i)$, 
\end{enumerate}
for any combination $i$, $j$ $(i<j)$, then $b_{ij}$ takes a positive value.  
\end{proposition}
\begin{figure}
\begin{center}
\begin{tikzpicture}[x=1mm, y=1mm]
  \draw[->] (0, 0) -- (120, 0)node[right]{$t$};
  \draw[-] (5, 2) -- (5, -2)node[left]{$0$};
  \draw (25, 0) to [out=60, in=120] (90, 0);
  \draw (25, 0)node[below] {$t_1$};
  \draw (90, 0)node[below] {$s_1$};
  \draw (30, 0) to [out=70, in=110] (80, 0);
  \draw (30, 0)node[below] {$s_5$};
  \draw (80, 0)node[below] {$t_5$};
  \draw (40, 0) to [out=80, in=100] (60, 0);
  \draw (65, 0) to [out=80, in=100] (75, 0);
  \draw (100, 0) to [out=80, in=100] (110, 0);
  \draw (40, 0)node[below] {$t_3$};
  \draw (65, 0)node[below] {$s_2$};
  \draw (100, 0)node[below] {$t_4$};
  \draw (60, 0)node[below] {$s_3$};
  \draw (75, 0)node[below] {$t_2$};
  \draw (110, 0)node[below] {$s_4$};
\end{tikzpicture}
\end{center}
\caption{An example of parameters $t_i$ and $s_i$.}
\label{diag1}
\end{figure}
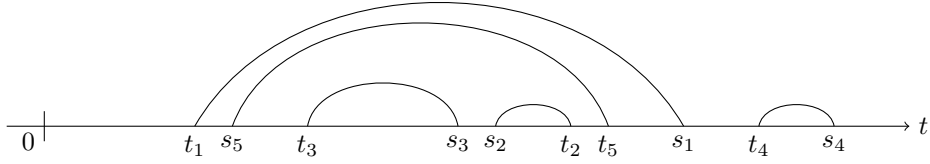
\textbf{Proof.} \ Figure \ref{diag1} shows an example of the parameters satisfying the conditions.  
The conditions in the proposition mean each line connected between $t_i$ and $s_i$ does not intersected with others.  The conditions derive $(t_i-t_j)(s_i-s_j)(t_i-s_j)(s_i-t_j)>0$ holds for any $i$, $j$ and it gives $b_{ij}>0$ from the definition.   \par
\begin{figure}  
  \begin{center}
   \begin{picture}(400, 120)
  \put(50,10){\includegraphics[width=6cm, height=3cm, clip]{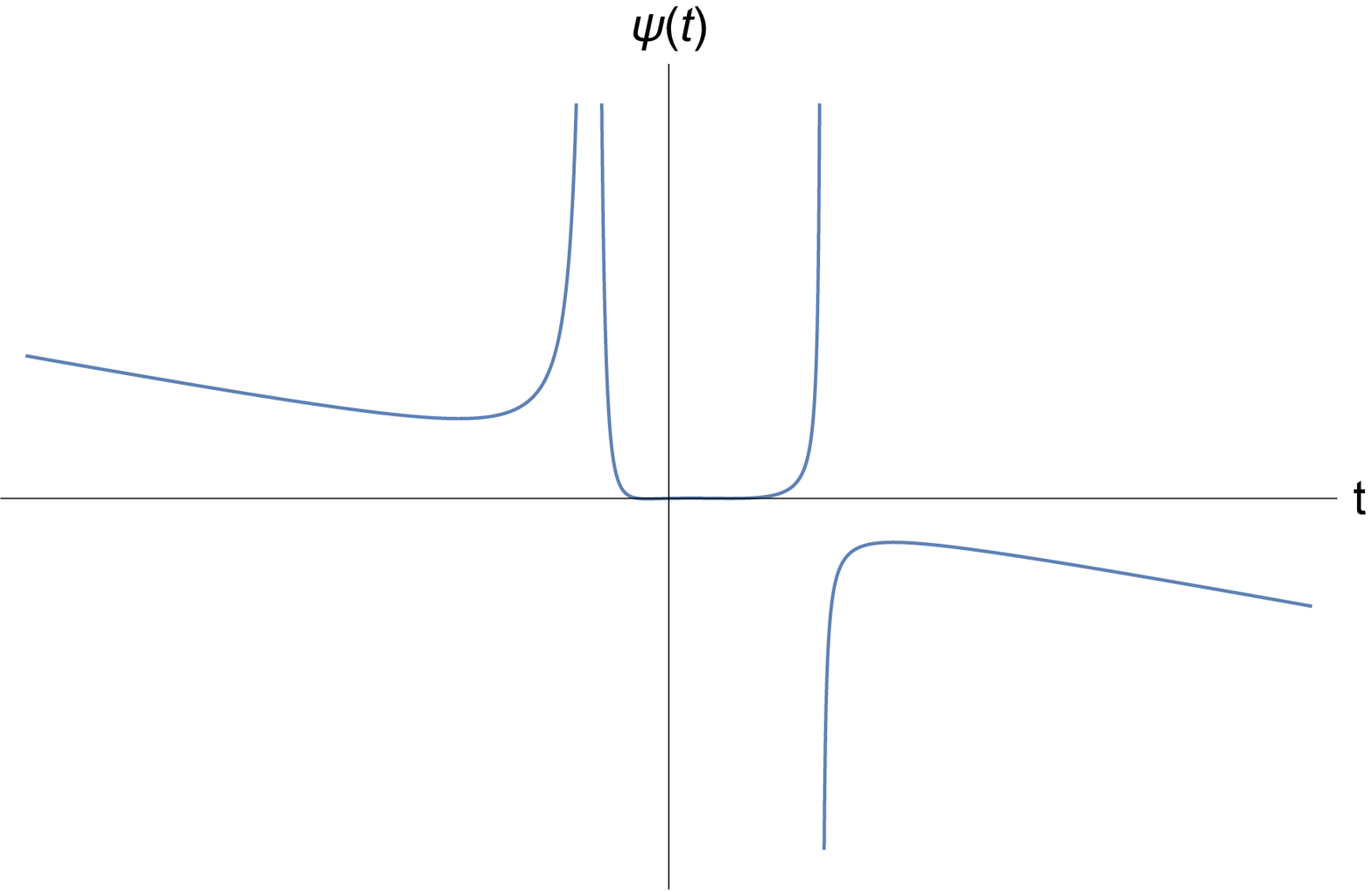}} 
  \put(250,10){\includegraphics[width=4cm, height=3cm, clip]{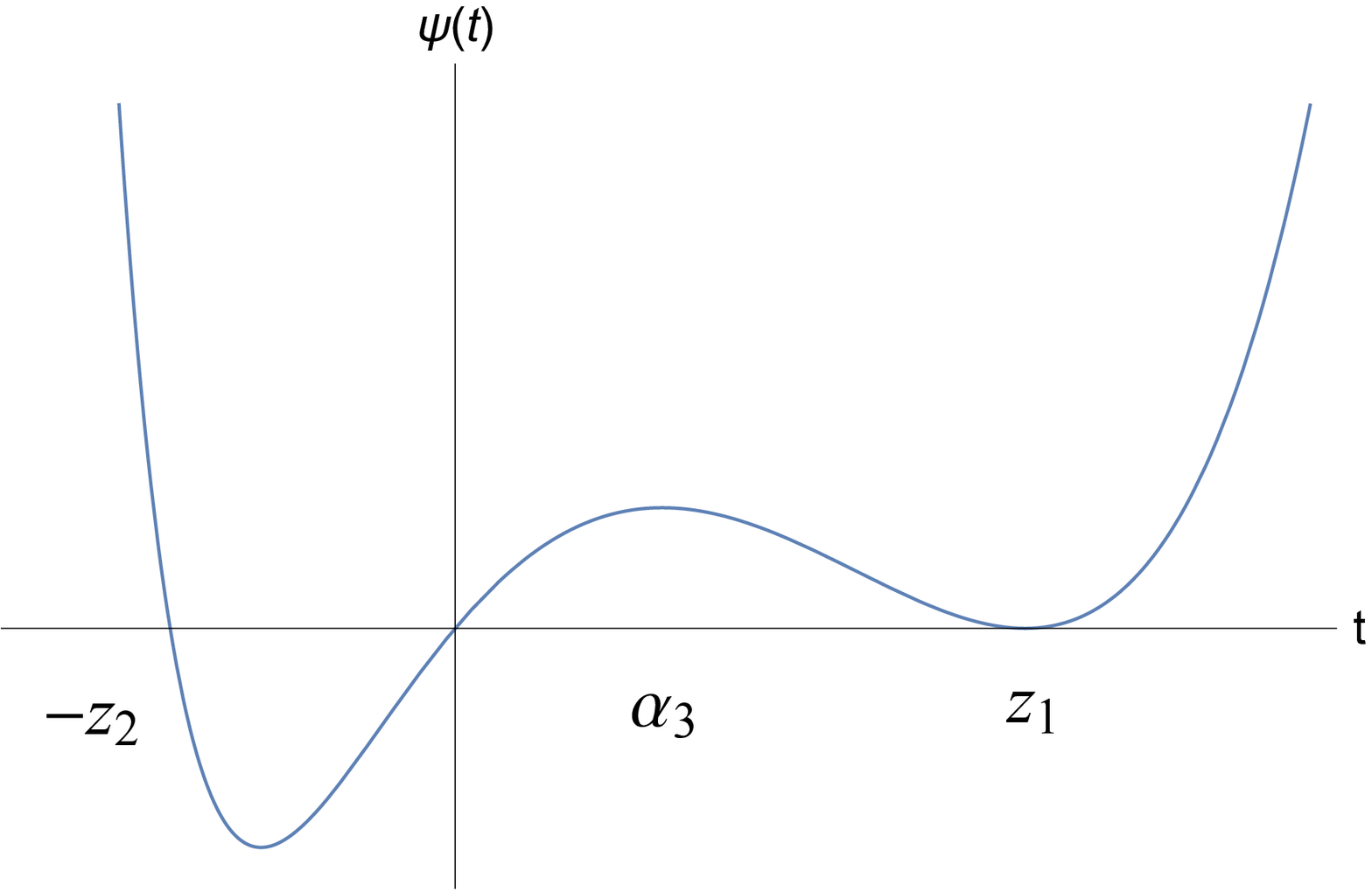}} 
   \end{picture}
\caption{Plot of $\psi(t)$. The right figure shows near the origin.  }
\label{fig1}
  \end{center}
\end{figure}
\begin{figure}  
  \begin{center}
   \begin{picture}(400, 120)
  \put(0,10){\includegraphics[width=4cm, height=3cm, clip]{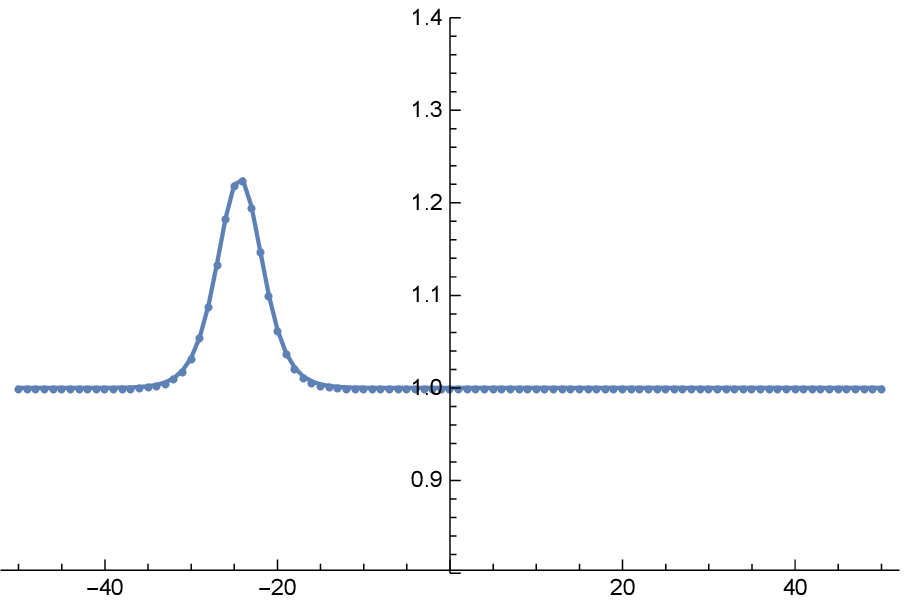}} 
  \put(150,10){\includegraphics[width=4cm, height=3cm, clip]{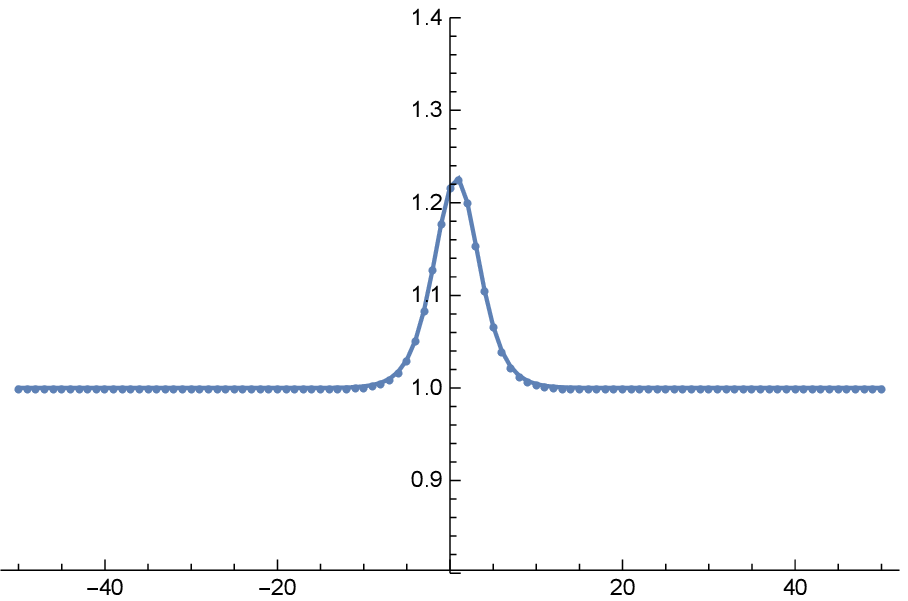}} 
  \put(300,10){\includegraphics[width=4cm, height=3cm, clip]{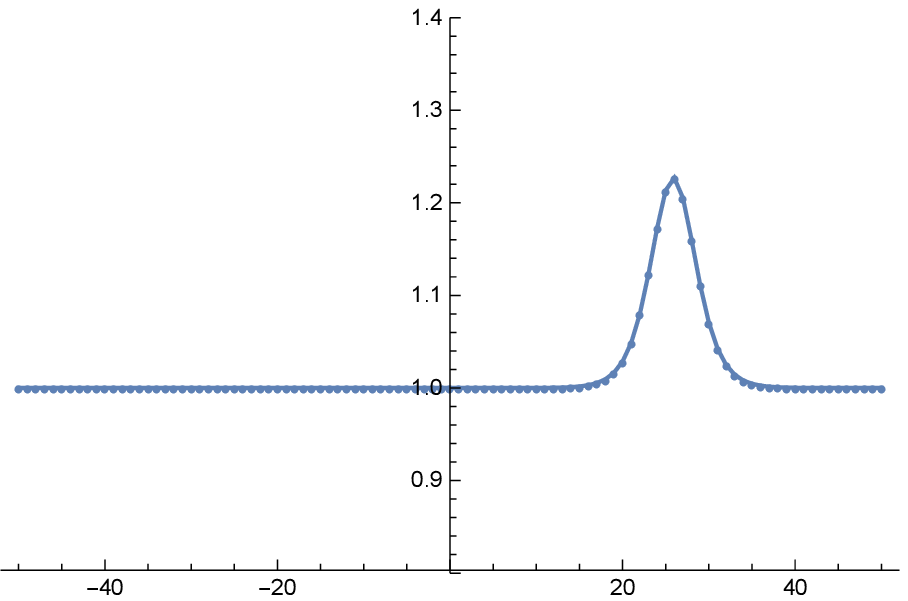}} 
  \put(35,0){\scriptsize $m=-10$} 
  \put(200,0){\scriptsize $m=0$} 
  \put(350,0){\scriptsize $m=10$} 
   \end{picture}
\caption{$1$-soliton solution $u^m_n$ with $t, s\in J_1$.  }
\label{fig2}
  \end{center}
\end{figure}
\begin{figure}  
  \begin{center}
   \begin{picture}(400, 120)
  \put(0,10){\includegraphics[width=4cm, height=3cm, clip]{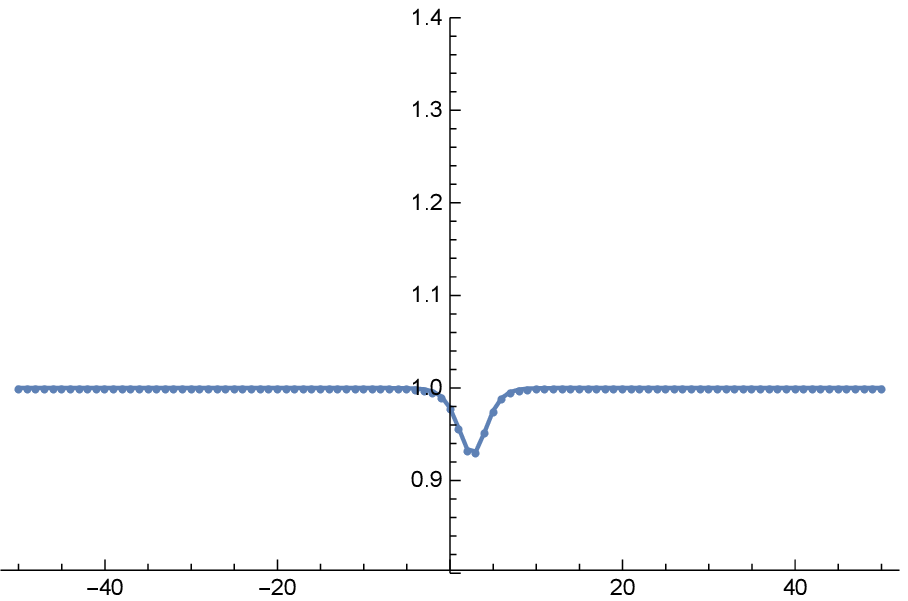}} 
  \put(150,10){\includegraphics[width=4cm, height=3cm, clip]{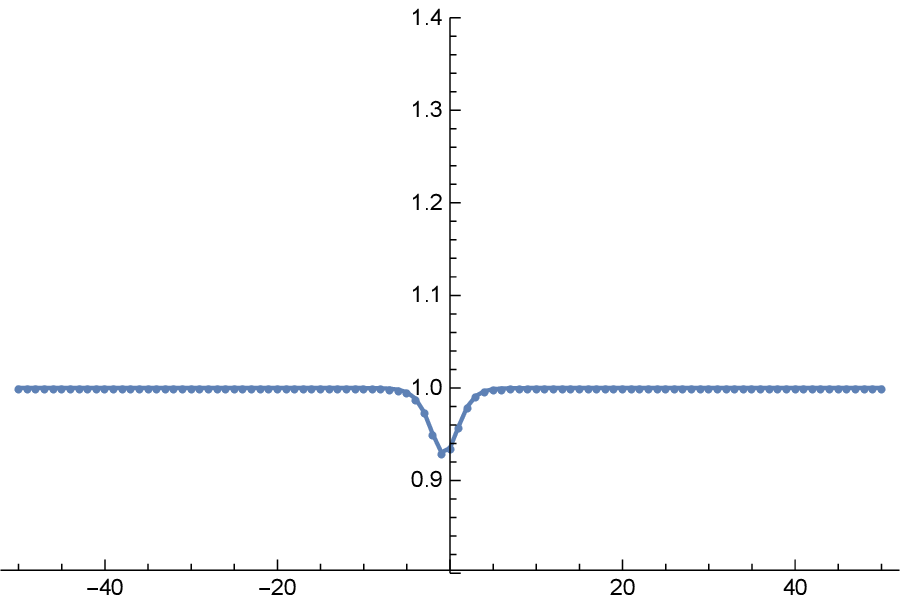}} 
  \put(300,10){\includegraphics[width=4cm, height=3cm, clip]{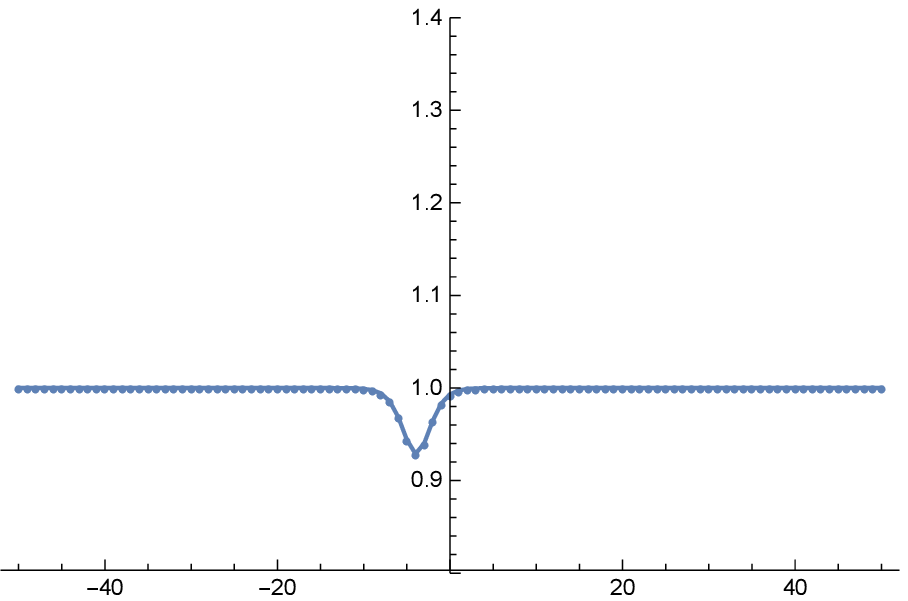}} 
  \put(35,0){\scriptsize $m=-10$} 
  \put(200,0){\scriptsize $m=0$} 
  \put(350,0){\scriptsize $m=10$} 
   \end{picture}
\caption{$1$-soliton solution $u^m_n$ with $t, s\in J_2$. }
\label{fig3}
  \end{center}
\end{figure}
\begin{figure}  
  \begin{center}
   \begin{picture}(400, 120)
  \put(0,10){\includegraphics[width=4cm, height=3cm, clip]{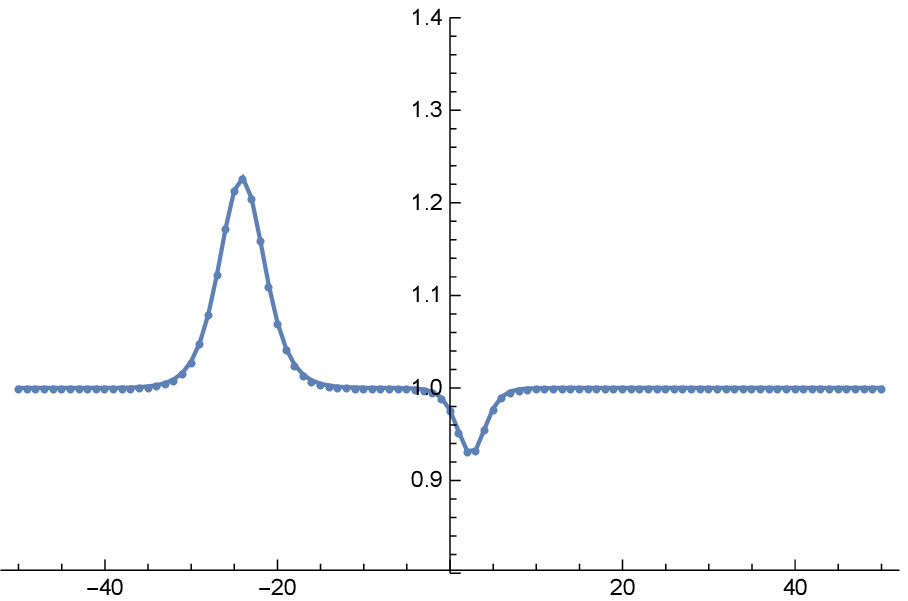}} 
  \put(150,10){\includegraphics[width=4cm, height=3cm, clip]{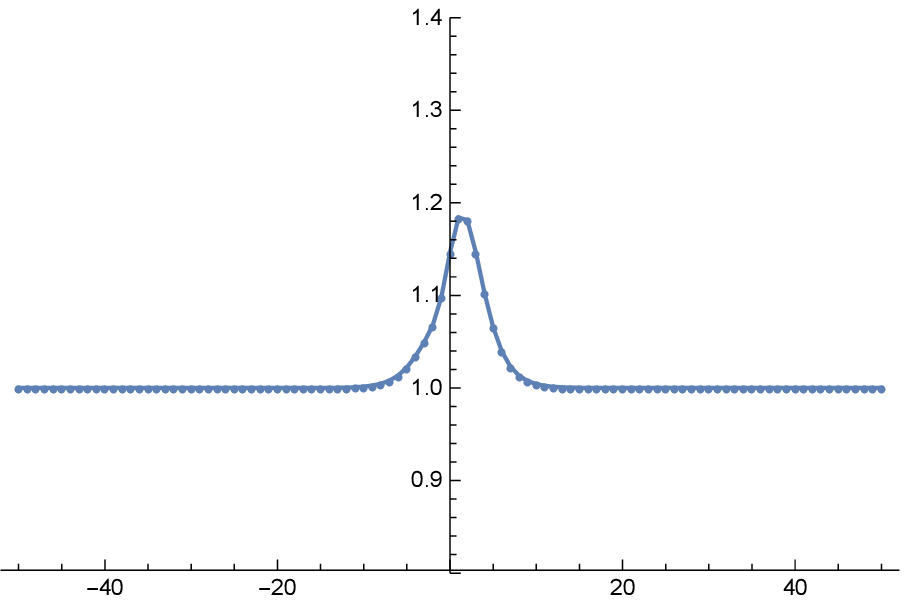}} 
  \put(300,10){\includegraphics[width=4cm, height=3cm, clip]{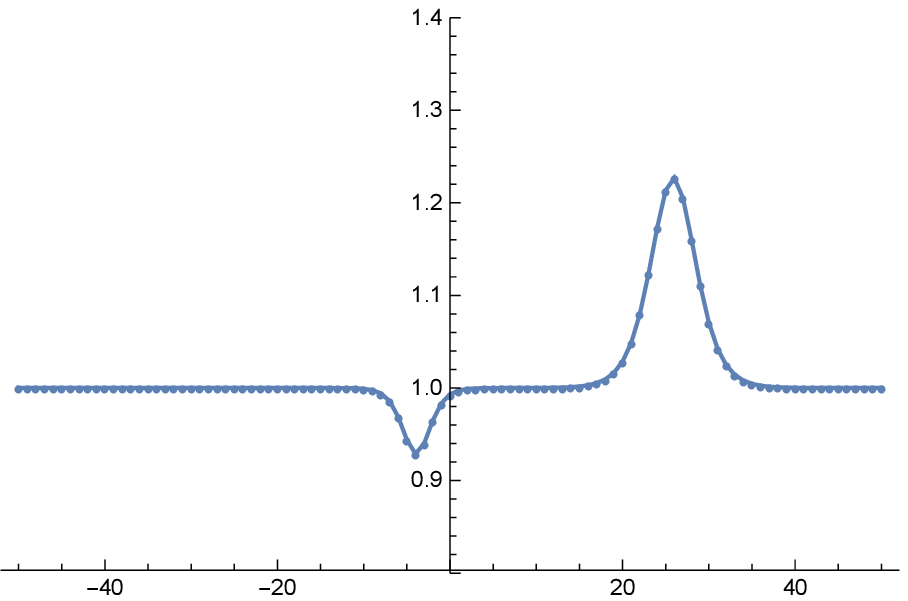}} 
  \put(35,0){\scriptsize $m=-10$} 
  \put(200,0){\scriptsize $m=0$} 
  \put(350,0){\scriptsize $m=10$} 
   \end{picture}
\caption{Mixed $(1+1)$-soliton solution $u^m_n$  with $t_1, s_1\in J_1$ and $t_2, s_2\in J_2$. }
\label{fig4}
  \end{center}
\end{figure}

These propositions hence can be summarized by the following theorem.   
\begin{theorem}  \label{theo:2-1}
  Suppose $c_i>0$ and parameters $t_i$ and $s_i$ satisfy the conditions given in Proposition \ref{prop:2-2} and \ref{prop:2-3} for $i=1, 2, \dots , N$, then the solution $\tau(p, q, r)$ take positive values for any $p, q, r\in \bf{Z}$.  
\end{theorem}
\indent Now let us derive the soliton solution of (\ref{eq:dmix}) under the assumptions given in Theorem \ref{theo:2-1}.  Hereafter we set $z_k$ as (\ref{def:z}).  By applying the transformations (\ref{trans}), the solution (\ref{Nsol0}) is transformed to 
\begin{equation}  \label{Nsol}
  \hat \tau(l, m, n)= \sum _{0\le k\le N} \sum _{I_k \subset [N] } \left( \prod _{i\in I_k} c_i\frac{\phi(t_i)}{\phi(s_i)}  \prod _{i, j\in I_k \atop{i<j }} b_{ij}\right),
\end{equation}
where 
\begin{equation}  \label{def:phi}
  \phi(t) = \left(\frac{a_3(t)}{(a_1(t))^2a_2(t)}\right)^l \left( \frac{a_3(t)}{a_1(t)}\right)^m (a_1(t))^n.  
\end{equation}
If $t_i$, $s_i$ satisfy the relation   
\begin{equation}  \label{cond of a}
  \frac{a_3(t_i)}{(a_1(t_i))^2a_2(t_i)}=\frac{a_3(s_i)}{(a_1(s_i))^2a_2(s_i)} 
\end{equation}
for $i=1, 2, \dots , N$, then the constraint condition $\hat \tau(l+1, m, n)=\hat \tau(l, m, n)$ holds, hence (\ref{Nsol}) satisfies (\ref{eq:dmix}).   
Considering the fluctuation of $\psi(t):= a_3(t)/a_1^2(t)/a_2(t)$ from elementary calculus, one can find the following(See Figure \ref{fig1}).  
\begin{itemize} 
  \item There exist three points $\alpha _i$ $(i=1, 2, 3)$ such that $\frac{d}{dt}\psi(\alpha _i)=0$ on $0<t$.  Each $\alpha _i$ satisfies $\alpha _1\in (\frac{z_1z_4}{z_3}, \infty)$, $\alpha _2=z_1$, $\alpha _3\in (0, z_1)$ respectively.  
  \item $\psi(t)$ is monotone increasing on $(z_1z_4/z_3, \alpha_1)$, $(z_1, z_1z_4/z_3)$ and $(0, \alpha _3)$.  
  \item $\psi(t)$ is monotone decreasing on $(\alpha _1, \infty)$ and $(\alpha _3, z_1)$.  
  \item $\psi(t)\ge 0$ on $(0, z_1z_4/z_3)$.  $\psi(t)<0$ on $(z_1z_4/z_3, \infty)$.  
\end{itemize}
Therefore if $t$ belongs to either $J_1=\left(\frac{z_1z_4}{z_3}, \infty \right)$ or $J_2= (0, z_1)$, except for $\alpha_1$, $\alpha_3$, then there exists a unique value $s$ in the same interval such that $s\not=t$ and satisfying (\ref{cond of a}).  In particular, these parameters $t_1, t_2, \dots , t_N, s_1, s_2, \dots s_N$ hold the conditions in Theorem \ref{theo:2-1}.  These results are summerized as the following theorem.  
\begin{theorem}
The solution of (\ref{eq:dmix}) is expressed by   
\begin{equation}
  f^m_n =  \sum _{0\le k\le N} \sum _{I_k \subset [N] } \left( \prod _{i\in I_k} c_i\frac{\phi(t_i)}{\phi(s_i)}  \prod _{i, j\in I_k \atop{i<j }} b_{ij}\right),
\end{equation}
where 
\begin{equation}  
  \phi(t) =  \left( \frac{a_3(t)}{a_1(t)}\right)^m (a_1(t))^n.  
\end{equation}
Here the parameters $t_i$, $s_i$ belong to the same intervals $J_1$ or $J_2$, and they satisfy the dispersion relation (\ref{cond of a}) for $i=1, 2, \dots , N$.  Moreover $f^m_n$ take positive values for any $m$, $n$ under $c_i>0$.   
\end{theorem}
Figure \ref{fig2} shows the one soliton solution with $t, s \in J_1$ and Fig. \ref{fig3} does with $t, s \in J_2$.  Here $m$ and $n$ denote time and space variables.  In Fig.\ref{fig2}, we set $t=3$, $(d_1, d_2)=(1/2, 1/3)$ and $s$ so that satisfying (\ref{cond of a}).  
The solitary wave moves to the right side.  In Fig.\ref{fig3}, we set $t=3/10$, $(d_1, d_2)=(1/2, 1/3)$ and $s$ satisfies (\ref{cond of a}).  The solitary wave moves to the left side slowly.  In each case, the velocity of solitary wave is given by $1-\log (a_3(t)/a_3(s))/\log(a_1(t)/a_1(s))$.  Figure \ref{fig4} shows the interaction of two different soliton solutions.  We set $(t_1, t_2)=(3, 3/10)$, $(d_1, d_2)=(1/2, 1/3)$, $s_1$ and $s_2$ satisfy (\ref{cond of a}) respectively.  \\
It is noted  distinct parameters $t$, $s$ in $J_2$ do not exist when $d_2=0$ since $\psi(t)$ is monotone decreasing on $J_2$.  It means the discrete KdV equation does not admit the solution with $t$, $s$ in $J_2$.  
\section{Ultradiscrete Soliton Solutions}
\subsection{One Soliton Solution}
In order to derive ultradiscrete solutions for (\ref{eBB}), we introduce new parameters for each $J_1$, $J_2$:
\begin{equation}
\begin{aligned}
   &t = t^{(1)} +\frac{z_1z_4}{z_3} \qquad \text{for $t\in J_1$}, \\
   &t = \frac{t^{(2)}z_1}{1+t^{(2)}} \qquad  \text{for $t\in J_2$}.  
\end{aligned}
\end{equation}
These replacements enable us to take $t^{(i)}$ for any positive values.  Then for $t\in J_1$, by using $z_1+z_2+z_3-z_4=0$, we have 
\begin{equation}
\begin{aligned}
  a_1(t) =& \frac{z_2z_4+z_3 t}{z_1-t } \\
  =& -\frac{z_3( z_4(z_1+z_2) +z_3t^{(1)} )}{z_1(z_1+z_2) + z_3 t^{(1)} } .  
\end{aligned}
\end{equation} 
We may neglect the signature since it can be cancelled with $a_1(s)$.  Thus, from (\ref{def:z}) and $d_1=e^{-\delta _1/\varepsilon }$, $d_2=e^{-\delta _2/\varepsilon }$, $t^{(i)}= e^{-T^{(i)}/\varepsilon }$, we obtain the ultradiscrete analogue of $|a_1(t)|$ as 
\begin{equation}
\begin{aligned}
  |a_1(t)| \to  & \max(-\delta_1, T^{(1)}) -\max (-2\delta_1 , T^{(1)}) \\
  =: & A_1(T^{(1)}).  
\end{aligned}
\end{equation} 
Similarly, we also obtain the ultradiscrete analogues of $|a_2(t)|$, $|a_3(t)|$ as follows.  
\begin{equation}
\begin{aligned}
  |a_2(t)| \to& A_2(T^{(1)}) = T^{(1)}-\max(T^{(1)}, -\delta _1), \\
  |a_3(t)| \to& A_3(T^{(1)}) = \max(T^{(1)}, -\delta _1).  
\end{aligned}
\end{equation} 
Hence we obtain the ultradiscrete analogue of the one soliton solution with $t, s\in J_1$.  
\begin{equation}  \label{sol:caseI}
\begin{aligned}
  F^m_n= \max(0,  & m(A_3(T^{(1)})-A_3(S^{(1)})-A_1(T^{(1)})+A_1(S^{(1)}) )\\
  & +n(A_1(T^{(1)})-A_1(S^{(1)}))+C),
\end{aligned}
\end{equation} 
where $T^{(1)}$, $S^{(1)}$ satisfy the dispersion relation, 
\begin{equation}
  A_3 (T^{(1)})-2A_1(T^{(1)})-A_2(T^{(1)})=A_3 (S^{(1)})-2A_1(S^{(1)})-A_2(S^{(1)}), 
\end{equation}
which is obtained by ultradiscretizing (\ref{cond of a}).  The above is simplified as 
\begin{equation}
  |T^{(1)}+2\delta _1|=|S^{(1)}+2\delta _1|.  
\end{equation} 
This leads $S^{(1)}= -T^{(1)}-4\delta _1$ due to $T^{(1)}\not=S^{(1)}$.  If we denote $\Omega $, $K$ as the coefficients of $m$, $n$: 
\begin{equation}  \label{def:OK}
\begin{aligned}
  \Omega =&  A_3(T^{(1)})-A_3(S^{(1)})-A_1(T^{(1)})+A_1(S^{(1)}),  \\
  K= &A_1(T^{(1)})-A_1(S^{(1)}) ,
\end{aligned}
\end{equation}
then $\Omega $ can take any value since $\Omega = T^{(1)}+2\delta_1$ and we find $K = \frac{1}{2}(|\Omega -\delta _1|-|\Omega +\delta _1|)$ holds.   Therefore, (\ref{sol:caseI}) is rewritten by  
\begin{equation}
  F^m_n= \max(0, \Omega m +K n +C).  
\end{equation}
This is the one soliton solution for (\ref{eBB}).  It is noted this solution corresponds to the one of the ultradiscrete KdV equation\cite{uKdV}.    \par 
For $t\in J_2$, we obtain 
\begin{equation}
\begin{aligned}
  |a_1(t)| \to& A_1(T^{(2)}) = \max(T^{(2)}-\delta _1, -\delta _2) +\delta _1, \\
  |a_2(t)| \to& A_2(T^{(2)}) = \max(T^{(2)}-2\delta _1, -\delta _1) -\max(T^{(2)}-\delta _1, -\delta _2),\\
  |a_3(t)| \to& A_3(T^{(2)}) = T^{(2)}-\delta _1-\max(0, T^{(2)}).  
\end{aligned}
\end{equation} 
The dispersion relation
\begin{equation}
 A_3 (T^{(2)})-2A_1(T^{(2)})-A_2(T^{(2)})=A_3 (S^{(2)})-2A_1(S^{(2)})-A_2(S^{(2)})
\end{equation}
holds by setting  
\begin{equation}
  S^{(2)} = -T^{(2)} +\delta_1-\delta _2 -\max(T^{(2)}-\delta _1, 0) -\frac{1}{2} \max(-T^{(2)}-\delta _2, 0).  
\end{equation}
If we denote $Q$, $P$ as the coefficients of $m$, $n$, 
\begin{equation} \label{def:QP}
\begin{aligned}
  Q =& A_3(T^{(2)})-A_3(S^{(2)})-A_1(T^{(2)})+A_1(S^{(2)}),   \\
  P= &A_1(T^{(2)})-A_1(S^{(2)}) , 
\end{aligned}
\end{equation}
then $P$ can take any value and $  Q= \max(0, P-\delta _2)+\min(0, P+\delta _2)$ holds.  Therefore, we obtain another one soliton solution for (\ref{eBB}).  
\begin{equation}
  F^m_n= \max(0, Q m +P n +C).  
\end{equation}
It is noted this solution corresponds to the one of the ultradiscrete Toda equation\cite{Matsukidaira}.   
\subsection{Mixed Soliton Solution}
\begin{figure}  
  \begin{center}
   \begin{picture}(400, 120)
  \put(-10, 105) {\vector(1,0){22}} 
  \put(-10, 105) {\vector(0,-1){22}} 
  \put(15, 103) {\small $n$} 
  \put(-13, 70) {\small $m$} 
  \put(0,10){\includegraphics[width=4cm, height=3cm, clip]{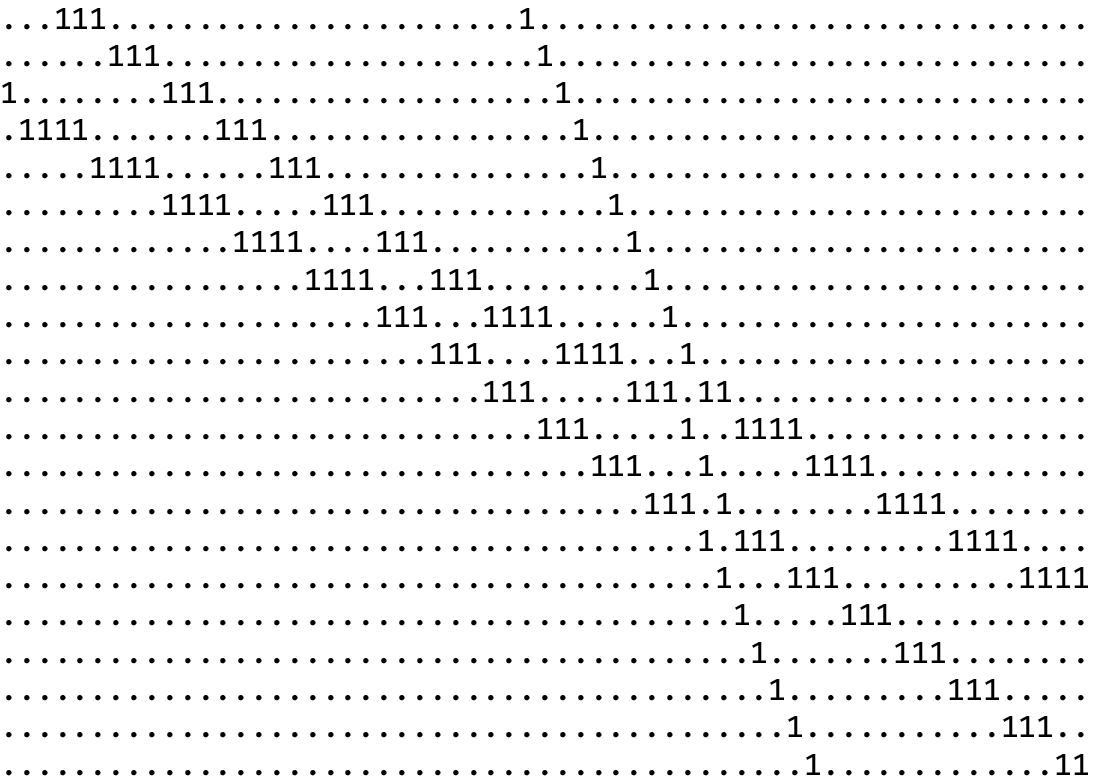}} 
  \put(130,10){\includegraphics[width=4cm, height=3cm, clip]{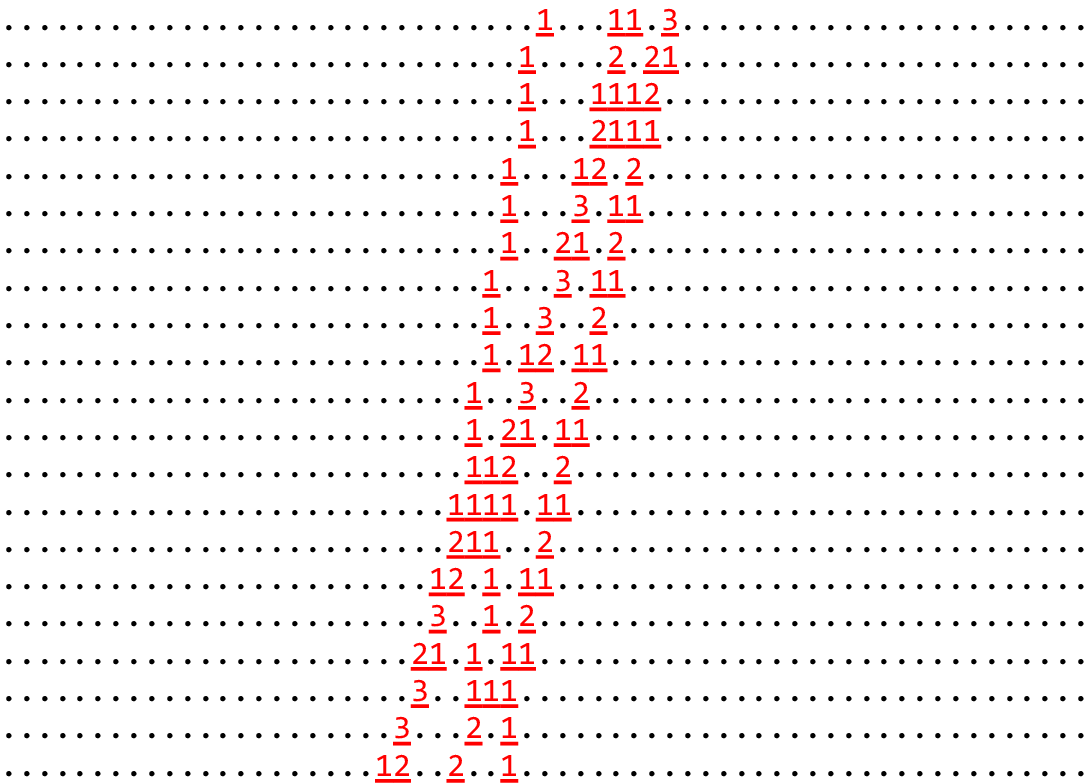}} 
  \put(260,10){\includegraphics[width=4cm, height=3cm, clip]{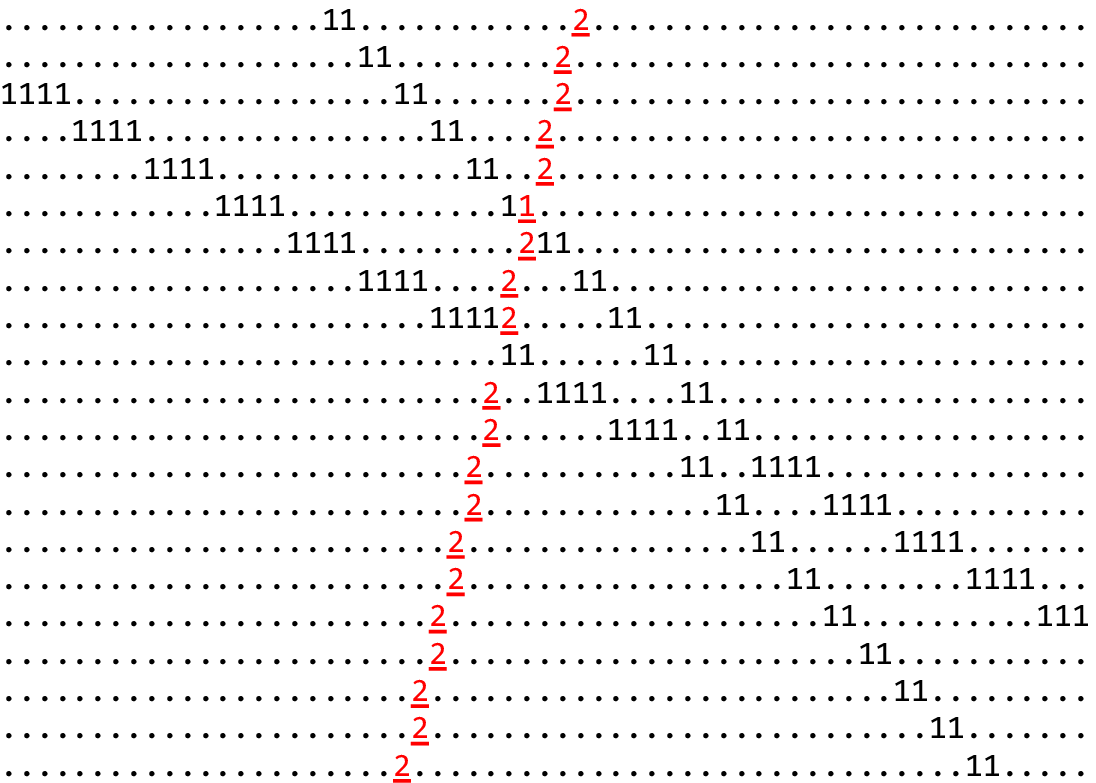}} 
   \end{picture}
\caption{Ultradiscrete soliton solutions $U^m_n$. The symbols $.$ and $\underline{1}$  denote $0$ and $-1$ respectively.  }
\label{fig6}
  \end{center}
\end{figure}

We consider the ultradiscrete analogue of $b_{ij}$.  We define 
\begin{equation}
\begin{aligned}
  &B(T_i^{(\alpha )}, T_j^{(\beta )})\\
   =& Cf (T^{(\alpha )}_i, T^{(\beta )}_j)+Cf (S^{(\alpha )}_i, S^{(\beta )}_j) -Cf (T^{(\alpha )}_i, S^{(\beta )}_j)-Cf (S^{(\alpha )}_i, T^{(\beta )}_j) 
\end{aligned}
\end{equation}
for $\alpha , \beta =1, 2$, where $Cf(T^{(\alpha )}_i, T^{(\beta )}_j)$ is the ultradiscrete analogue of $|cf (t_i, t_j)|$, in which $t_i$ and $t_j$ are replaced as $t_i^{(\alpha )}$, $t_j^{(\beta )}$.  For example, $\alpha =1$, $\beta =2$, we have
\begin{equation}
\begin{aligned}
  cf(t^{(1)}, t^{(2)})  =& \frac{1}{z_3}\frac{z_3t^{(2)}t^{(1)} + z_3t^{(1)} +z_1t^{(2)}(z_1+z_2)+z_1z_4 }{ (z_3t^{(1)}+z_1z_4)z_1t^{(2)} +(1+t^{(2)})z_1z_2z_4}  \\
 \to  & \max(T^{(1)}, T^{(1)}+T^{(2)}, T^{(2)}-2\delta _1, -\delta _1) \\
  &-\max(T^{(1)}+T^{(2)}, T^{(2)}-\delta _1, -\delta _2)+\delta _1  \\
  =:& Cf(T^{(1)}, T^{(2)}).  
\end{aligned}
\end{equation}
Notice $Cf(S^{(1)}, S^{(2)})$ can be expressed by $T^{(1)}$ and $T^{(2)}$ through the dispersion relations.
We can calculate  
\begin{equation}
\begin{aligned}
  B(T_i^{(1)}, T_j^{(1)}) =& 2\max(\min(\Omega_i, -\Omega_j), \min(-\Omega_i, \Omega_j)), \\
  B(T_i^{(2)}, T_j^{(2)}) =& \max(\min(-2P_i-Q_i, 2P_j+Q_j), \min(2P_i+Q_i, -2P_j-Q_j)), 
\end{aligned}
\end{equation}
and 
\begin{equation}
\begin{aligned}
  B(T_i^{(1)}, T_j^{(2)})   =& 
\begin{cases}
    -\max(\min(\Omega_i +2K_i, 2Q_j), \min(\Omega_i +2K_i-P_j+\delta _2, 0 )) &\quad (\Omega_i >0, P_j>0) \\
    \max(\min(\Omega_i +2K_i, -2Q_j), \min(\Omega_i +2K_i+P_j+\delta _2, 0 )) &\quad (\Omega_i >0, P_j<0) \\
    \max(\min(-\Omega_i -2K_i, 2Q_j), \min(-\Omega_i -2K_i-P_j+\delta _2, 0 )) &\quad (\Omega_i <0, P_j>0) \\
    -\max(\min(-\Omega_i -2K_i, -2Q_j), \min(-\Omega_i -2K_i+P_j+\delta _2, 0 )) &\quad (\Omega_i <0, P_j<0) 
\end{cases}
\end{aligned}
\end{equation}
where $\Omega_i$, $K_i$, $P_i$, $Q_i$ are defined by (\ref{def:OK}) and (\ref{def:QP}).  Due to $|b_{ij}|=|b_{ji}|$, $B(T_i^{(2)}, T_j^{(1)})$ can be also obtained.  Thus the following holds.  
\begin{theorem}
The solution of (\ref{eBB}) is expressed by
\begin{equation}  
  F^m_n =  \max _{0\le k\le N+M} \max _{I_k \subset [N+M] }  \left( \sum _{i\in I_k} \Phi_i(m, n) + \sum _{i, j\in I_k \atop{i<j }} B_{ij}\right), 
\end{equation}
where $N$, $M$ are non negative integers. When $k=0$, we define $\max _{I_0\subset [N+M] }$ is $0$. Function $\Phi_i(m, n)$ is defined by 
\begin{equation}
  \Phi_i(m, n)= 
\begin{cases}
  \Omega_i m +  K_i n +C_i& \qquad (i=1,2, \dots , N) \\
   Q_i m +P_in +C_i & \qquad (i=N+1,N+2, \dots , N+M) 
\end{cases}
\end{equation}
\begin{equation}
  K_i= \frac{1}{2}\left( |\Omega_i -\delta _1| - |\Omega_i +\delta _1|\right) , \qquad  Q_i= \max(0, P_i-\delta _2)+\min (0, P_i+ \delta _2), 
\end{equation}
with arbitrary parameters $P_i$, $\Omega _i$ and $C_i$.  The interaction factor $B_{ij}$ is defined by 
\begin{equation}
\begin{aligned}
  B_{ij}=&
\begin{cases}
  2\max(\min(\Omega_i, -\Omega_j), \min(-\Omega_i, \Omega_j))  &\quad  (i<j\le N )\\
  \max(\min(-2P_i-Q_i, 2P_j+Q_j), \min(2P_i+Q_i, -2P_j-Q_j)) &\quad  (N< i<j )\\
  B'_{ij}  & \quad (i\le N <j)
\end{cases}, 
\end{aligned}
\end{equation}
where 
\begin{equation}
\begin{aligned}
  B'_{ij}  =&
\begin{cases}
    -\max(\min(\Omega_i +2K_i, 2Q_j), \min(\Omega_i +2K_i-P_j+\delta _2, 0 )) &\quad (\Omega_i >0, P_j>0) \\
    \max(\min(\Omega_i +2K_i, -2Q_j), \min(\Omega_i +2K_i+P_j+\delta _2, 0 )) &\quad (\Omega_i >0, P_j<0) \\
    \max(\min(-\Omega_i -2K_i, 2Q_j), \min(-\Omega_i -2K_i-P_j+\delta _2, 0 )) &\quad (\Omega_i <0, P_j>0) \\
    -\max(\min(-\Omega_i -2K_i, -2Q_j), \min(-\Omega_i -2K_i+P_j+\delta _2, 0 )) &\quad (\Omega_i <0, P_j<0) 
\end{cases}.  
\end{aligned}
\end{equation}
\end{theorem}
We note the solution in case $M=0$ corresponds to the soliton solution of the ultradiscrete KdV equation.  In other words,  this solution $U^m_n$ behaves as the original Box-Ball system through the transformations
\begin{equation}  \label{def:ultrauvx}
\begin{aligned}
  U_n^m=&F_{n+1}^m+F_n^{m+1}-F_n^m-F_{n+1}^{m+1}, \\
  V^m_n =& F_{n}^{m+1}+F_n^{m-1}-2F_{n}^{m}, \\
  X^m_n =& F_{n+1}^{m}+F_{n-1}^{m}-2F_{n}^{m},  
\end{aligned}
\end{equation}
which are the ultradiscrete analogues of (\ref{def:uvx}).  In fact, we can prove $-\delta_2 + X^m_n +  X^m_{n+1} < 0$ holds for any $m$, $n$ when $M=0$\cite{NagaiBack}.  Figure \ref{fig6} shows the behaviour of the ultradiscrete soliton solution $U^m_n$.   
The left figure in Fig.\ref{fig6} shows $(3+0)$ soliton solution with $(\Omega_1, \Omega_2, \Omega_3)=(1, 3, 4)$ and $(\delta _1, \delta _2)=(1, 2)$.  As mentioned in Section 2, its evolution obeys the original Box-Ball system with the box capacity $\delta _1$.  The middle figure in Fig.\ref{fig6} shows $(0+3)$ soliton solution with $(P_1, P_2, P_3 )=(3 ,4 ,5)$.  The notation $\underline {1}$ denotes $-1$. We can observe $U^m_n$ take negative values and move to the left side as discrete solutions.   The right figure in Fig.\ref{fig6} shows $(2+1)$-mixed soliton solution with $(\Omega_1, \Omega_2, P_1)=(2, 4, 4)$.  We can observe they pass through each other without losing their identities.  

\section{Continuous limit}

In this Section, we investigate the continuous limit of the equation (\ref{eq:dmix}).
Let us start with rewriting the equation (\ref{eq:dmix}) using the Hirota's D-derivative.
\begin{eqnarray}
\left\{- (1 + d_1) \cosh\left(- D_m - \frac{1}{2} D_n\right) 
+ d_1 \cosh\left(D_m - \frac{1}{2} D_n \right) \right. \nonumber \\ 
\left.
+ d_2 \cosh\left(\frac{3}{2} D_n\right)
+ (1-d_2) \cosh\left(\frac{1}{2} D_n \right)
\right\} f \cdot f = 0
\label{eq:dmixD1}
\end{eqnarray}
Here, $D_m$, $D_n$ are the Hirota's D-derivative corresponding to the 
variables $m$,$n$ respectively.

To take the continuous limit we introduce the new variables $x_1$, $x_5$ 
and relate them with the original variables as 
\begin{eqnarray*}
D_m = 2 \epsilon D_1 + \frac{2}{3} a_2 \epsilon^3 D_1 + 
\frac{2}{5} a_3  \epsilon^5 D_5, \\
D_n = 2k \epsilon D_1 + \frac{2}{3} b_2 \epsilon^3 D_1 
+ \frac{2}{5} b_3 \epsilon^5 D_5.  
\end{eqnarray*}
Here, $D_1,\ D_5$ are the Hirota's D-derivative of the variables 
$x_1,\ x_5$, and $k,\ a_2,\ a_3,\ b_2,\ b_3$ are arbitrary complex parameters.
$\epsilon$ is the real parameter which represents the lattice spacing of 
the discrete variables.
Furthermore, we replace the parameters $d_1,\ d_2$ as
\begin{eqnarray*}
d_1 = - \frac{1}{2} + \frac{-k + b_2 \epsilon^2}{4k^2},\ 
d_2 = \frac{a_2 \epsilon^2 + 1}{4k^2}.  
\end{eqnarray*}
Then equation (\ref{eq:dmixD1}) becomes as follows
\begin{eqnarray*}
\left\{
(k(2k-1)+b_2 \epsilon^2) \cosh\left(-D_m - \frac{1}{2} D_n\right)
+ \left(k(2k+1) - b_2 \epsilon^2 \right) \cosh\left(D_m - \frac{1}{2} 
   D_n \right) \right. \\  \left.
+(-1 -a_2 \epsilon^2) \cosh\left(\frac{3}{2} D_n\right)
+(1 - 4k^2 + a_2 \epsilon^2) \cosh\left(\frac{1}{2} D_n \right)
\right\} f \cdot f = 0
\end{eqnarray*}
Taking the small $\epsilon$ limit we obtain the following equation 
as the coefficient of the sixth order term of the parameter $\epsilon$.
\begin{eqnarray}
&& \left\{
108 k (a_3 k - b_3) D_1 D_5
- 8k^2 (k^2-1)(4k^2-1) D_1^6 \right. \nonumber \\ 
&& \quad \left. \quad -40 k (k^2 - 1) (a_2k + 2 b_2) D_1^4
+40(a_2 k - b_2)^2 D_1^2
\right\} f \cdot f  = 0
\label{eq:cmixD1}
\end{eqnarray}
This equation contains the terms of $D_1 D_5 f \cdot f,\ D_1^6 f \cdot f,\ 
D_1^4 f\cdot f$ and $D_1^2 f \cdot f$.
As a special case, let us assume the following form of the parameters 
$a_i,\ b_i$. 
\begin{eqnarray*}
&&a_2 = c \left( \frac{-2 \sqrt{(k^2 - 1) (4k^2 - 1)} - 4k^2 + 1}{3}\right), \\
&&b_2 = c k \left( \frac{\sqrt{(k^2-1)(4k^2-1)} - 4k^2 + 1}{3}\right), \\
&&a_3 = - \frac{2}{3} + \frac{10}{3}k^2 - \frac{8}{3}k^4,\\
&&b_3 = 0
\end{eqnarray*}
Substituting these expressions into the equations (\ref{eq:cmixD1}), 
we obtain the following equation.
\begin{eqnarray*}
- \frac{3k^2 (k^2 - 1)(4k^2 - 1)}{45} 
(9 D_1 D_5 + D_1^6 - 5c D_1^4 -  5c^2 D_1^2) f \cdot f = 0
\end{eqnarray*}
This is nothing but the bilinear form of the KdV-Sawada-Kotera equation 
appearing in the paper \cite{BilinearKSK}.
\section{Concluding Remarks}
We have derived an soliton equation and its solution from the generalized discrete BKP equation.  
The equation admits two types of solitary wave solutions. 
One is the solution which moves to the positive direction 
and has the positive amplitude while the other does the negative direction and amplitude.  
The type of the solution depends on whether the parameter $t$ belongs to $J_1$ or $J_2$.  
Interestingly while these solutions have the same form and dispersion relation, 
their behaviours are quite different.  We also have derived its ultradiscrete analogues.  
The dispersion relations in the ultradiscrete soliton solutions have the same expressions 
of the ultradiscrete KdV and ultradiscrete Toda equation.  Furthermore we have shown the relation between our equation and the KdV-Sawada-Kotera equation through the continuous limit.    
\section*{Acknowledgement}
The authors are grateful to Prof. Ralpf Willox for valuable advices.  
\renewcommand{\theequation}{A.\arabic{equation}}
\setcounter{equation}{0}
\appendix 
\def\thesection{Appendix \Alph{section}.}
\section{Transformation}
We shall derive (\ref{Nsol0}) from 
\begin{equation}  \label{gen sol}
  \bar \tau(p, q, r) =\sum _{k_1=1}^2\sum _{k_2=1}^2\dots \sum _{k_N=1}^2 \prod_{1\le i<j\le N} cf(t_i^{(k_j)}, t_j^{(k_i)}) \prod _{i=1}^N c_i^{(k_i)} \varphi (t_i^{(k_i)})  , 
\end{equation}
which is a soliton solution of (\ref{eq:gdBKP}) given in \cite{Shinzawa}.  Here $t_i^{(k_i)}$ and $c_i^{(k_i)}$ are arbitrary parameters, $\varphi (t)$ is defined by  
\begin{equation}
  \varphi(t) = (a_1(t))^p (a_2(t))^q ( a_3(t) )^r .  
\end{equation}
Using the gauge transformation, that is, by multiplying (\ref{gen sol}) with 
\begin{equation*}
  \frac{1}{\prod_{1\le i<j\le N }cf(t_i^{(2)}, t_j^{(2)})\prod _{i=1} ^Nc_i^{(2)}\varphi (t_i^{(2)})}, 
\end{equation*}
we have another form of soliton solution,
\begin{equation}  \label{2-2}
  \tau(p, q, r)  =\sum _{k_1=1}^2\sum _{k_2=1}^2\dots \sum _{k_N=1}^2 \prod_{1\le i<j\le N} \frac{cf(t_i^{(k_j)}, t_j^{(k_i)})}{cf(t_i^{(2)}, t_j^{(2)}) } \prod _{i=1}^N \frac{c_i^{(k_i)} \varphi (t_i^{(k_i)})}{c_i^{(2)} \varphi (t_i^{(2)})}.  
\end{equation}
The terms with  
\begin{equation*}
  k_i=
\begin{cases}
1 \qquad  &  (i\in I:=\{i_1, i_2,\dots , i_l \}) \\
2 \qquad  &  (i\in I^c:=[N]- \{i_1, i_2,\dots , i_l \}) 
\end{cases}  
\end{equation*}
in (\ref{2-2}) are expressed by 
\begin{equation} \label{2-3} 
  \prod _{i\in I }\frac{c_i^{(1)}}{c_i^{(2)}}\frac{\varphi(t_i^{(1)})}{\varphi(t_i^{(2)})}\left(  \prod_{1\le i<j\le N \atop i, j\in I} \frac{cf(t_i^{(1)}, t_j^{(1)})}{cf(t_i^{(2)}, t_j^{(2)}) }\prod_{1\le i<j\le N \atop i\in I , j\in I^c} \frac{cf(t_i^{(2)}, t_j^{(1)})}{cf(t_i^{(2)}, t_j^{(2)}) }\prod_{1\le i<j\le N \atop i\in I^c, j\in I} \frac{cf(t_i^{(1)}, t_j^{(2)})}{cf(t_i^{(2)}, t_j^{(2)}) } \right).
\end{equation}
If we replace free parameters $c_i$ as 
\begin{equation*}
  \frac{c_i^{(1)}}{c_i^{(2)}} \to c_i\prod _{j=1, i\not=j} ^N \frac{ cf(t_i^{(2)}, t_j^{(2)})}{ cf(t_i^{(2)}, t_j^{(1)})}, 
\end{equation*}
then (\ref{2-3}) is reduced to 
\begin{equation} 
  \prod _{i\in I }c_i\frac{\varphi(t_i^{(1)})}{\varphi(t_i^{(2)})}\prod _{i, j\in I \atop{i<j }} b_{ij}  
\end{equation}
since $cf(t, s)=-cf(s, t)$ and 
\begin{equation*}
   \prod _{i\in I \atop{j\not =i }}b_{ij} =  \prod _{i, j\in I \atop{i<j }}b_{ij} \prod _{i, j\in I \atop{i<j }}b_{ji} \prod _{i\in I, j \in I^c \atop{i<j }}b_{ij}\prod _{j\in I, i \in I^c \atop{i<j }} b_{ji}  
\end{equation*}
hold.  Therefore (\ref{2-2}) corresponds to (\ref{Nsol0}).  

\end{document}